\documentclass[aps,prl,reprint,twocolumn,showpacs,superscriptaddress,amsmath,amssymb]{revtex4-1}

\usepackage{graphicx}  
\usepackage{dcolumn}   
\usepackage{bm}        
\usepackage{hyperref}
\usepackage{verbatim}  
\usepackage{natbib}
\usepackage{epsfig}
\usepackage{epstopdf}
\usepackage{booktabs}  
\usepackage{color}
\usepackage[toc,page]{appendix}
\usepackage{placeins}
\usepackage{float}
\usepackage{multirow}
\usepackage[normalem]{ulem}
\usepackage{amsmath}
\usepackage{tabu}
\usepackage[utf8]{inputenc}
\usepackage{rotating}
\usepackage{multirow}
\usepackage[export]{adjustbox}

\begin{document}

\title{Drop Dynamics on Liquid Infused Surfaces: The Role of the Lubricant Ridge}

\author{Muhammad S. Sadullah}
\affiliation{Department of Physics, Durham University, Durham, DH1 3LE, UK}
\author{Ciro Semprebon}
\affiliation{Department of Mathematics, Physics and Electrical Engineering, 
Northumbria University, Newcastle upon Tyne NE1 8ST, UK}
\author{Halim Kusumaatmaja}
\email{halim.kusumaatmaja@durham.ac.uk}
\affiliation{Department of Physics, Durham University, Durham, DH1 3LE, UK}
\date{\today}

\begin{abstract}
	We employ a free energy lattice Boltzmann method to study the dynamics 
	of a ternary fluid system consisting of a liquid drop driven by a body force across 
	a regularly textured substrate, infused by a lubricating liquid. 
	We focus on the case of partial wetting lubricants 
	and observe a rich interplay between contact line pinning and viscous 
	dissipation at the lubricant ridge, which become dominant at large and small
	apparent angles respectively. Our numerical investigations further demonstrate 
	that the relative importance of viscous dissipation at the lubricant ridge depends on
	the drop to lubricant viscosity ratio, as well as on the shape of the wetting ridge.
\end{abstract}

\maketitle

\section{Introduction}

\begin{figure}[tb]
	\centering
	\includegraphics{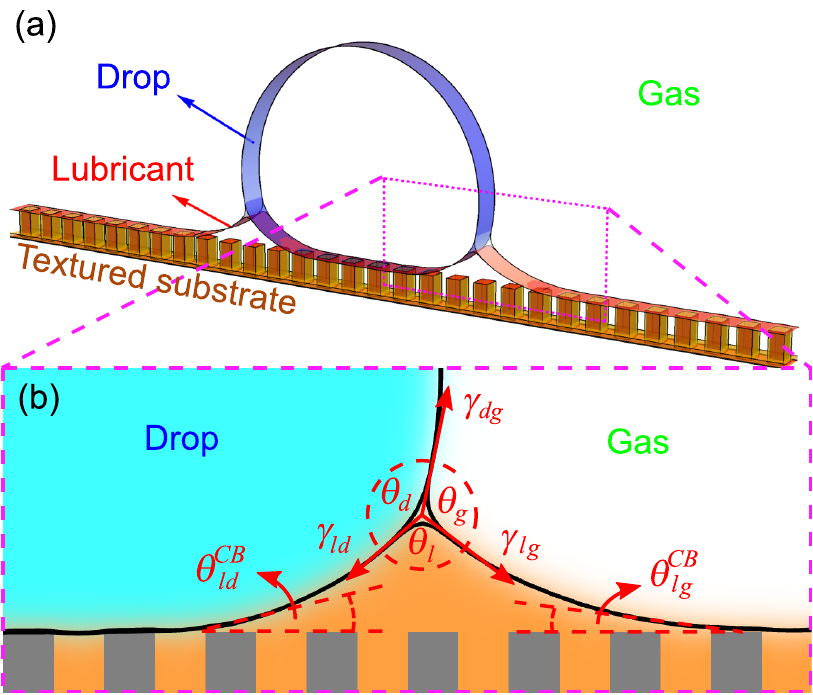}
	\caption{(a) Rendering of a quasi 3D setup of an LIS system where a drop is sitting on a 
	textured substrate infused with a lubricant. 
	(b) Magnification of the lubricant ridge.
	$\gamma_{dg}$, $\gamma_{lg}$, and $\gamma_{ld}$ are the drop-gas, lubricant-gas, 
	and drop-lubricant surface tensions; $\theta_{d}$, $\theta_{g}$, and $\theta_{l}$ 
	are the Neumann angles of the drop, gas and lubricant; 
	$\theta_{ld}^{CB}$ and $\theta_{lg}^{CB}$ are the drop-lubricant and lubricant-gas 
	contact angles assuming a Cassie-Baxter approximation.}
	\label{fgr1:lis_illustration}
\end{figure}

Liquid Infused Surfaces (LIS) are liquid repellent surfaces constructed by infusing 
a lubricant into textured substrates \cite{Quere2005}, as illustrated in Fig.~\ref{fgr1:lis_illustration}. 
Drops placed on LIS move very easily under small perturbations and will shed away at a small tilting angle, 
regardless of their surface tensions \cite{Wong2011}. These surfaces can also be designed 
to withstand high pressure and self-heal from physical damages \cite{Wong2011}, 
which distinguish them from other liquid repellent surfaces such as superhydrophobic surfaces \cite{Ma2006}.

LIS are relatively easy to fabricate. The primary requirements are a rough 
solid substrate with strong affinity toward the lubricant, and the drop needs to be immiscible 
to the lubricant \cite{Kim2013}. These advantageous features have given rise to many potential 
industrial applications, such as to reduce energy consumption in fluid transports \cite{Shirtcliffe2009}, 
to simplify cleaning and maintenance processes \cite{Lu2015}, to prevent damage due to fouling \cite{Ferrari2015}, 
and to annihilate product leftover for smart liquid packaging \cite{liquiglide}.
For many of these applications, efficient and effective control of the drop dynamics on LIS 
is required, yet to date such control remains poorly understood. 

Compared to the more commonly studied cases of smooth and superhydrophobic surfaces 
\cite{Richard1999,Mahadevan1999,hodges2004,Mognetti2010,Thampi2013,Moradi2011}, the main distinguishing 
feature of LIS is the presence of the infusing lubricant, forming a ridge as shown in Fig.~\ref{fgr1:lis_illustration}.
Thus the central aim of this work is to shed light on the role of the lubricant ridge 
in the dynamics of drops on LIS.

Based on thermodynamic arguments, 
\citeauthor{Smith2013} showed that a liquid drop placed on LIS may invade the corrugation 
and replace the infusing lubricant, or it can sit on top of the corrugation with 
the lubricant present underneath the drop \cite{Smith2013}. If the lubricant is perfectly wetting 
the substrate, the drop and the corrugated surface is separated by a thin film, 
and no pinning of the contact lines take place. However, closer inspection employing 
confocal microscopy revealed that this case is unlikely for a number of common 
lubricants, as they form in contact to the solid with a small but finite contact angle 
\cite{Smith2013, Schellenberger2015}. As such, on one hand, the surface roughness helps 
to contain the lubricant; on the other hand, it is also the source of contact line pinning 
and contact angle hyeteresis.

The presence of lubricant meniscus also introduces competing dissipation mechanisms
acting on a drop as it moves across LIS. For example, \citeauthor{Keiser2017}
have highlighted that viscous dissipation may occur predominantly in the drop
or in the lubricant depending on the ratio between the drop and lubricant viscosities \cite{Keiser2017}.
However, most studies to date consider only drops with apparent contact angles close to $90^\circ$ 
\cite{Smith2013, Keiser2017}, and the impact of the shape of the lubricant meniscus on drop 
mobility remain  unexplored. To cover such gaps, here we will investigate these variations 
systematically using the lattice Boltzmann simulation method. 
In particular, we focus on the interplay between the contact line pinning induced by the surface 
corrugation and the viscous dissipation in the lubricant and drop phases. 


\section{Numerical Method}
\label{method}
\subsection{Ternary free-energy lattice-Boltzmann method}
To simulate liquid drops on LIS, we employ a ternary fluid model able to account 
for three bulk fluids (drop, gas and lubricant), their fluid-fluid interfacial tensions, 
and the fluid interactions with a solid substrate. The free energy is given by 
\cite{Kim2007,Semprebon2016a,Dong2018}
\begin{eqnarray}
F &=& \sum_{m=1}^3  \int_\Omega \left[\frac{\kappa_m}{2} C_m^2 (1 - C_m)^2+ \frac{\alpha^2 \kappa'_m}{2} (\nabla C_m)^2 \right] \, \mathrm{d}V \nonumber\\
&-& \sum_{m=1}^3  \int_{\partial\Omega} h_m C_m \, \mathrm{d}S.
\label{eq:ternary_free_energy}
\end{eqnarray}

By construction the first term in Eq.~\ref{eq:ternary_free_energy} corresponds 
to a double well potential. Taken separately, each double well potential has 
minima at concentrations $C_m = 0$ (fluid component $m$ is absent) and 
$C_m = 1$ (fluid component $m$ is present). In our simulations, we initialise the 
system such that $\sum_{m=1}^3 C_m =1$ at any point in the simulation box, 
with three physically meaningful bulk states corresponding to the drop 
(${\bf C} = [1,0,0]$), the gas (${\bf C} = [0,1,0]$) 
and the lubricant (${\bf C} = [0,0,1]$) phases.

The second term in Eq.~\ref{eq:ternary_free_energy} is related to the energy 
penalty at an interface between two fluid phases. The interfacial tension between 
fluid phases $m$ and $n$, $\gamma_{mn}$ ($m,n=1,2,3$ and $m \neq n$), can be tuned 
by the $\kappa_m$ parameters via \cite{Semprebon2016a}
\begin{equation}
	\gamma_{mn} =\frac{\alpha}{6}(\kappa_m + \kappa_n),
	\label{eq:LB_surface_tension}
\end{equation}
where $\alpha = \sqrt{(\kappa'_m + \kappa'_n)/(\kappa_m + \kappa_n)}$ 
is a parameter we can tune to vary the interface width. 
Typically we choose $\alpha = 1$ in our simulations.

The $h_m$ parameters in the last term of Eq.~\ref{eq:ternary_free_energy} allow 
us to quantify the fluid-solid surface energies and correspondingly the contact angle 
of fluid $m$ on a solid surface in the presence of fluid $n$, $\theta_{mn}$.
The contact angle $\theta_{mn}$ is given by \cite{Semprebon2016a} 
\begin{equation}
	\cos \theta_{mn} = \frac{\gamma_{sn}-\gamma_{sm}}{\gamma_{mn}},
\end{equation}
where each solid-liquid tension $\gamma_{sm}$ include contribution from
both majority and minority phases, expressed by the integrals $I$ and $J$ respectively
\begin{eqnarray}
\gamma_{sm} &=& I_m + \sum_{n \neq m} J_n,\nonumber\\
I_m &=& \frac{\alpha k_m}{12}
-\frac{h_m}{2}
-\frac{4 h_m + k_m\alpha}{12}\sqrt{1+\frac{4h_m}{\alpha k_m}},\nonumber\\
J_n &=& \frac{\alpha k_n}{12}
-\frac{h_n}{2}
+\frac{4 h_n - k_n\alpha}{12}\sqrt{1-\frac{4h_n}{\alpha k_n}}.\nonumber
\end{eqnarray}

For ternary fluid systems in contact with an ideal flat substrate
only two out of the three contact angles are independent. 
For example, if $\theta_{12}$ and $\theta_{32}$ are specified, the remaining 
contact angle, $\theta_{31}$, is determined by the Girifalco-Good relation \cite{Girifalco1957}
\begin{equation}
	\cos \theta_{31} = \frac{\gamma_{32} \cos \theta_{32} - \gamma_{12} \cos \theta_{12}}{\gamma_{31}}.
	\label{eq:GoodGirifalco_relation}
\end{equation}
Any choice of the $h_m$ parameters fulfills Eq.~\ref{eq:GoodGirifalco_relation}.

In our approach, we apply variable transformations from $C_1$, $C_2$ and $C_3$ to 
three equivalent order parameters, $\rho = C_1 + C_2 + C_3$, $\phi = C_1 - C_2$, 
and $\psi = C_3$. For simplicity, here we have set the density $\rho = 1$ everywhere. 
This ``equal density" approximation is suitable for small Reynolds number ($Re$), 
which is the case in this work. At large $Re$, inertia becomes important, and the density 
ratios between the fluid components must be taken into account \cite{Wohrwag2018}. 
In terms of these order parameters, the equations of motion of the system are described 
by the continuity, Navier-Stokes and two Cahn-Hilliard equations
\begin{eqnarray}
		& \partial_t \rho + \vec{\nabla} \cdot \left( \rho \vec{v} \right) = 0, \label{eq:continuity}  \\
		& \partial_t (\rho \vec{v}) + \vec{\nabla} \cdot \left( \rho \vec{v} 
		\otimes \vec{v} \right) = - \vec{\nabla} \cdot \bf{P} + \vec{\nabla} \cdot 
		\left[ \eta ( \vec{\nabla v} +  \vec{\nabla v^T} ) \right], \,\,\,\,\,\,\, \label{eq:NSE}  \\
		& \partial_t  \phi + \vec{\nabla} \cdot (\phi \vec{v}) = M_\phi \nabla^2 \mu_\phi, \label{eq:CHEphi} \\
		&\partial_t \psi + \vec{\nabla} \cdot (\psi \vec{v}) = M_\psi \nabla^2 \mu_\psi, \label{eq:CHEpsi}
\end{eqnarray} 
where $\vec{v}$ is the fluid velocity, and $\eta$ is the fluid viscosity that generally depends 
on the local order parameters $\phi$ and $\psi$. The latter allows us to set different viscosities 
for the drop, lubricant, and gas components. The thermodynamic properties of the system, 
described in the free energy model in Eq.~\ref{eq:ternary_free_energy}, enter the equations 
of motion via the chemical potentials, $\mu_q = \delta F/ \delta q$, ($q = \phi$ and $\psi$), 
and the pressure tensor, $\bf{P}$, defined by
$\partial_\beta P_{\alpha\beta} = \phi\partial_\alpha \mu_\phi + \psi\partial_\alpha \mu_\psi$.
To solve the equations of motion, Eq.~\ref{eq:continuity} -~\ref{eq:CHEpsi}, we employ 
the ternary lattice Boltzmann algorithm described in Ref.~\citenum{Semprebon2016a}. 
More general details on the lattice Boltzmann method, including how it recovers the continuum 
equations of motion, can be found in Ref.~\citenum{Kusumaatmaja2010,Succi2001,HalimTim2017}.

\subsection{Simulation setup}

The majority of simulations are performed in a quasi three-dimensional simulation box, 
as shown in Fig.~\ref{fgr1:lis_illustration}. 
The dimension of the simulation box is $400 \times 10 \times 150$ LB units with the top surface 
bounded by a flat wall. The bottom solid surface is textured 
with a row of square posts of height $h = 10$ LB units, width $w = 5$ LB units, and periodicity $p = 10$ LB units. 
A periodic boundary condition is applied in the other two directions.

This quasi three-dimensional setup has the advantage of reducing the computational cost when compared to a full 3D simulation, 
while capturing the key 3D features. In the case of LIS, it preserves the essential feature of allowing 
the lubricant to flow in between the surface texture underneath the liquid drop. This setup has been 
successfully employed to study drop dynamics on flat and superhydrophobic surfaces
\cite{Kusumaatmaja2010,Mognetti2010,Moradi2010}.

The lubricant phase is initialised to fill the space between the posts and an additional layer of two 
lattice nodes on top of them, in order to allow the formation of a lubricant ridge at the two sides of the drop. 
To make sure the lubricant imbibes the bottom surface, the lubricant-drop contact angle 
$\theta_{ld}$ and the lubricant-gas contact angle $\theta_{lg}$ have to be smaller than the
critical angle $\theta_c$ for hemi-wicking. From thermodynamic considerations it can be shown that
$\cos \theta_c = (1-\phi_s)/(r-\phi_s)$, where $\phi_s$ and $r$ are respectively the solid fraction and roughness 
factor of the surface pattern \cite{Quere2008,Semprebon2014}. The texture employed in this work 
gives $\phi_s = 0.25$ and $r = 3$, which leads to $\theta_c \approx 74^\circ$. 

A hemispherical drop with radius $R = 60$ LB units is placed on top of the posts and is then allowed 
to reach equilibrium before a body force is introduced to mobilise the drops. Unless stated otherwise, 
we set the horizontal and the vertical components of body force to be equal, such that $G_z=-G_x$. 
This choice corresponds to an experimental setup where the substrate is tilted at an angle of 45$^\circ$. 
Adding a downward body force ensures the drop to remain attached to the substrate, especially when 
it has a large apparent angle. We find the steady state velocity of the drops to be insensitive 
to the value of $G_z$ as long as the drop size is smaller than the capillary length, 
$R < l_c = \sqrt{\gamma_{dg}/\rho |G_z|}$. 
To characterise the drop mobility, we will take advantage of two dimensionless parameters, 
the Bond number $Bo = R^2G_x/\gamma_{dg}$ and the capillary number $Ca = \eta_d V_x/\gamma_{dg}$, 
where $\gamma_{dg}$, $\eta_d$, and $V_x$ are the drop-gas surface tension, drop viscosity and 
drop velocity parallel to the solid surface.

\section{Drop morphologies in mechanical equilibrium}
\label{ApparentAngle}
In this section we will demonstrate that our ternary lattice Boltzmann approach can accurately simulate
drop morphologies in mechanical equilibrium on LIS.
For a liquid drop placed on an ideal smooth surface, the material contact angle, $\theta_{dg}^Y$, 
is given by the Young's law, which arises from the force balance between the interfacial tensions 
at the three-phase contact line:
\begin{equation}
\cos \theta_{dg}^Y = \frac{\gamma_{sg} - \gamma_{sd}}{\gamma_{dg}},
\label{eq:young_law}
\end{equation} 
where $\gamma_{sg}$, $\gamma_{sd}$, and $\gamma_{dg}$ are the solid-gas, solid-drop and drop-gas 
interfacial tensions respectively. Here we employ the superscript $Y$ to distinguish the material 
contact angle from the effective contact angle under the Cassie-Baxter approximation (superscript $CB$).

\begin{figure}
	\centering
	\includegraphics{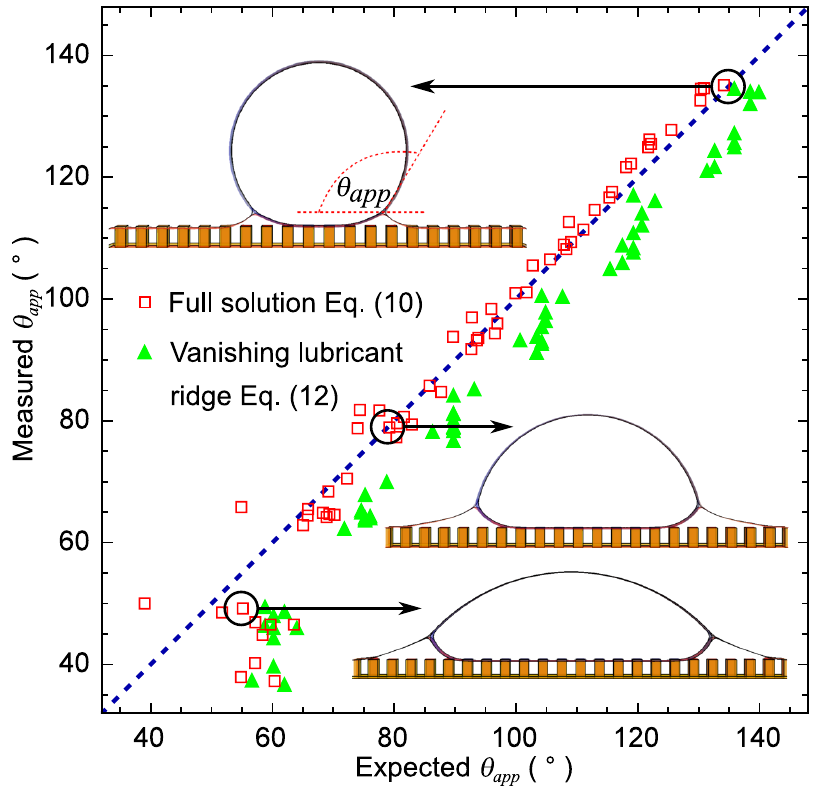}
	\caption{Comparison between $\theta_{app}$ obtained from our simulations against the 
	predicted values from both the full solution (Eq.~\ref{eq:fullsolution}) and the vanishing 
	lubricant ridge approximation (Eq.~\ref{eq:vanishingmeniscus}). The top left inset 
	illustrates how $\theta_{app}$ is measured at the drop-lubricant-gas triple line.}
	\label{fgr2:apparent_angle}
\end{figure}

For a drop placed on LIS, the solid-gas-drop contact line does not exist, and thus Eq.~\ref{eq:young_law} 
does not represent a physically meaningful condition. In contrast there exist three alternative three-phase 
lines (see Fig.~\ref{fgr1:lis_illustration}(b)): drop-lubricant-gas, drop-lubricant-solid, gas-lubricant-solid. 
To characterise how much the drop spreads on LIS, it is useful to introduce the notion of 
an apparent contact angle. As illustrated in Fig.~\ref{fgr2:apparent_angle} (top left), the apparent angle can be 
defined with respect to the horizontal plane at the drop-lubricant-gas triple line. In the limit of 
small but finite lubricant ridge, we have recently shown that the apparent angle need to satisfy the following relation \cite{Semprebon2016b}:
\begin{equation}
\frac{\sin \theta_g [\cos \theta_{ld}^{CB} - \cos (\theta_d - \theta_{app})]}
{\sin \theta_d [\cos \theta_{lg}^{CB} - \cos (\theta_{app} + \theta_g)]}
=\bigg(1-\frac{\Delta P_{dg}}{\Delta P_{lg}}\bigg).
\label{eq:fullsolution}
\end{equation}
Here $\theta_{\alpha\beta}^{CB}$ is the averaged wettability expressed by the
Cassie-Baxter contact angle \cite{Quere2008},
\begin{equation}
\cos \theta_{\alpha\beta}^{CB} = \phi_s \cos \theta_{\alpha\beta}^Y + (1-\phi_s),
\label{eq:cassie_baxter}
\end{equation}
which accounts for the fact that the drop and gas phases lie on top of a composite solid-lubricant interface. 
The quantity $\Delta P_{dg} / \Delta P_{lg}$ is the ratio between the Laplace pressures at the drop-gas and 
lubricant-gas interfaces. Since the Laplace pressure is given by 
$\Delta P_{\alpha\beta}=2\gamma_{\alpha\beta} / R_{\alpha\beta}$, 
where $R_{\alpha\beta}$ is the mean radius of curvature for the $\alpha\beta$ interface, 
$\Delta P_{dg} / \Delta P_{lg}$ is directly related to the size ratio between the lubricant ridge and the drop. 
In the strict limit of vanishing lubricant ridge, $\Delta P_{dg}/\Delta P_{lg} \rightarrow 0$, Eq.~\ref{eq:fullsolution} 
can be simplified to 
\begin{equation}
\cos\theta_{app}=\frac{\gamma_{lg}}{\gamma_{dg}}\cos \theta_{lg}^{CB} - \frac{\gamma_{ld}}{\gamma_{dg}}\cos \theta_{ld}^{CB}.
\label{eq:vanishingmeniscus}
\end{equation}
The main advantage of Eq.~\ref{eq:vanishingmeniscus} is that all variables on the 
right hand side are material parameters which can be measured independently. 
In contrast, the value of $\Delta P_{dg}/\Delta P_{lg}$ in Eq.~\ref{eq:fullsolution} is usually not known a priori.
However, it can be inferred from analysing the shape of the lubricant ridge.

In Fig.~\ref{fgr2:apparent_angle}, we compare the apparent angle, $\theta_{app}$, measured from our LB simulations 
once mechanical equilibrium is reached, against both the full solution in Eq.~\ref{eq:fullsolution} and the 
vanishing lubricant ridge approximation in Eq.~\ref{eq:vanishingmeniscus}. The range of apparent angles are 
obtained by varying the surface tensions $\gamma_{lg}$, $\gamma_{dg}$ and $\gamma_{ld}$, as well as the 
lubricant's material contact  
angles $\theta^Y_{lg}$ and $\theta^Y_{ld}$. For comparison against the full solution (Eq. \ref{eq:fullsolution}), 
we compute $\Delta P_{dg}/\Delta P_{lg}$ by measuring the radii of curvature of the drop-gas 
and lubricant-gas interfaces once mechanical equilibrium is reached in our simulations.
 
The measured apparent contact angles in our LB simulations are in very good agreement with the full solution.
When compared against the vanishing lubricant ridge approximation (Eq. \ref{eq:vanishingmeniscus}), 
the measured values of the apparent angle has a systematic deviation by several degrees. 
This deviation is expected since the size of the lubricant ridge in our simulations is not 
negligible compared to the drop size. Nonetheless, Eq.~\ref{eq:vanishingmeniscus} remains a good 
first estimate for predicting the apparent angle of drops on liquid infused surfaces, and the 
accuracy improves the smaller the lubricant meniscus is compared to the drop size.

\section{Translational Drop Mobility}
\label{translation}

\subsection{Variation in the ratio between drop and lubricant viscosities}

Recent experiment by \citeauthor{Keiser2017} suggests that there is a crossover between 
bulk drop and lubricant ridge dominated dissipation regimes, as the drop to lubricant 
viscosity ratio is varied \cite{Keiser2017}.
Here we aim to reproduce this crossover behaviour to demonstrate that our LB simulation 
can correctly capture the dynamics of drops moving across LIS.

\begin{figure}
	\centering
	\includegraphics{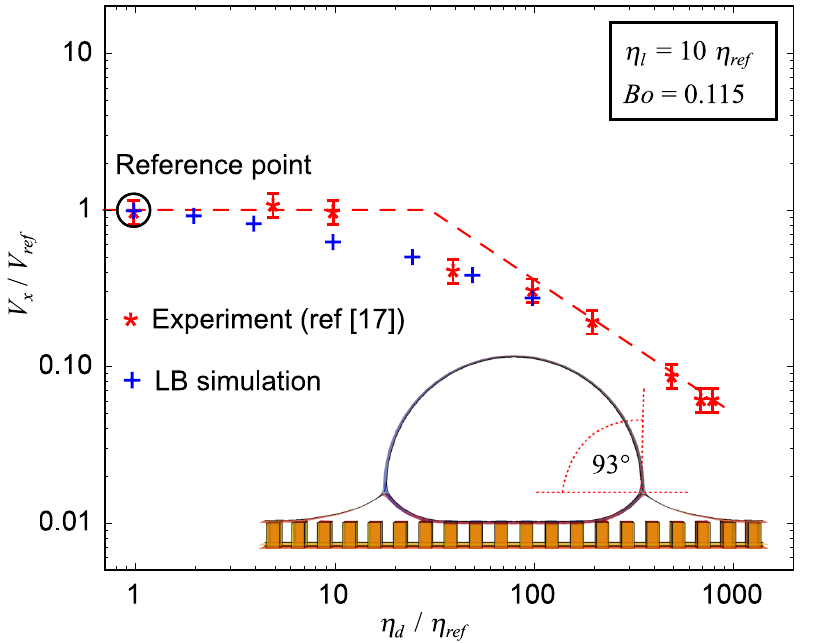}
		\caption{Comparison between our simulation results against experimental data by 
		\citeauthor{Keiser2017} \cite{Keiser2017}. Here the lubricant viscosity is fixed 
		at $\eta_l = 10 \, \eta_{ref}$, while the drop viscosity is varied. The reference 
		viscosity $\eta_{ref}$ in the experiment is water viscosity (1 mPa.s). To ensure 
		correct viscosity ratio between the drop and the air phases, we set 
		$\eta_{ref} = 50 \, \eta_g$ in our simulations. $V_{ref}$ is drop velocity when $\eta_d = \eta_{ref}$.}
		\label{fgr3:viscous_dissipation}
\end{figure}

We introduce a surface patterning, surface tensions, and a body force such that 
$\phi_s = 0.25$, $\theta_{app} = 93^\circ$, and $Bo$ = 0.115 to mimic the experimental 
setup in \cite{Keiser2017} ($\phi_s = 0.23$, $\theta_{app} = 90^\circ$, and $Bo$ = 0.115). 
The time averaged velocities of the drop's centre of mass from our simulations are reported 
by the blue plus symbols in Fig.~\ref{fgr3:viscous_dissipation}. The viscosity of water (about 50 
times larger than the viscosity of air), $\eta_{ref} = 50\eta_g$ =  1 mPa.s, is taken as 
the reference viscosity. We have also scaled the drop velocity by $V_{ref}$, taken to be 
the drop velocity $V_x$ when the drop viscosity is $\eta_d=\eta_{ref}$. For comparison, 
the experimental data from \citeauthor{Keiser2017} \cite{Keiser2017} are shown as red 
asterisks in Fig.~\ref{fgr3:viscous_dissipation}. 

\begin{figure*}
	\centering
	\includegraphics[width=\textwidth,keepaspectratio]{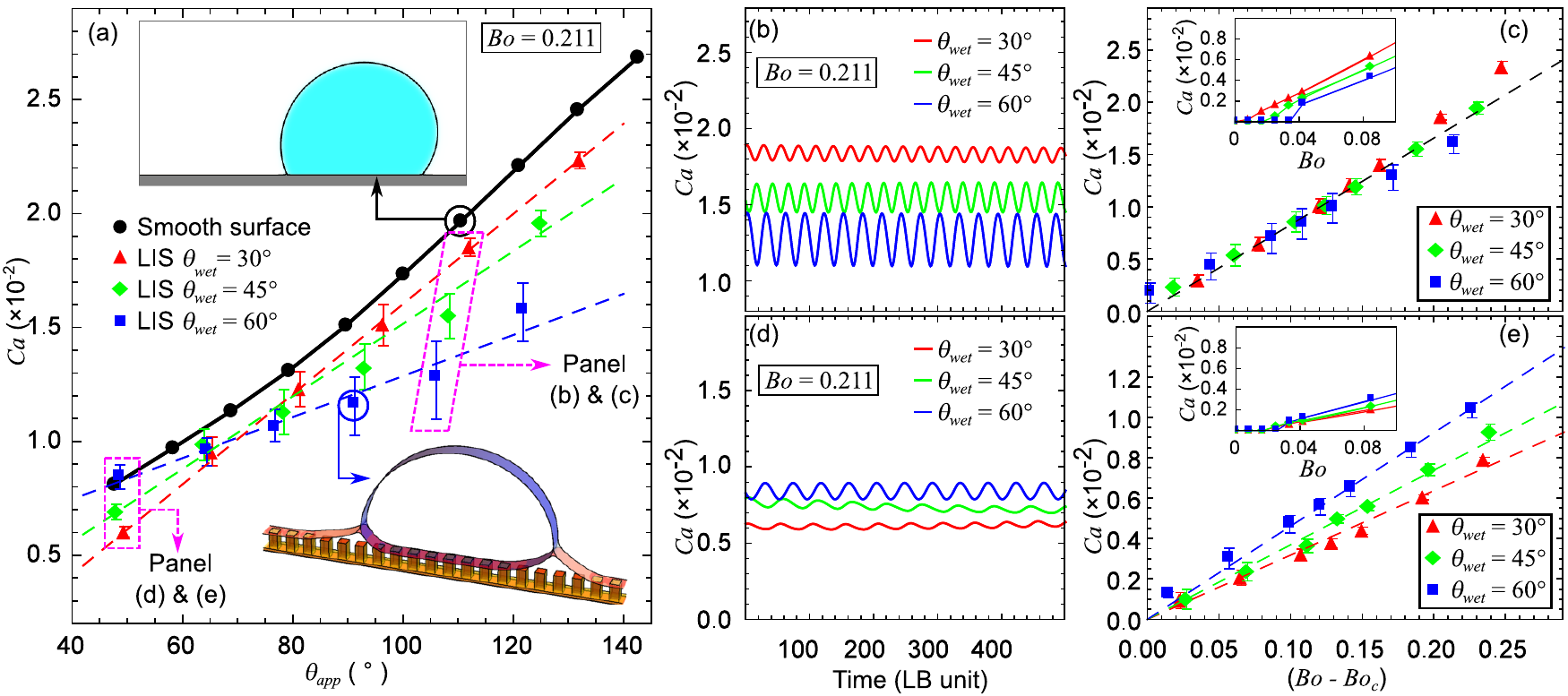}
	\caption{(a) Drop mobility on smooth surface (black dots) and LIS (red triangles, 
		green diamonds, and blue squares) plotted against $\theta_{app}$. $\theta_{wet}$ 
		is the wetting angle of the lubricant phase ($\theta^Y_{lg}=\theta^Y_{ld}=\theta_{wet}$).  
		The drop mobility is represented by the capillary number $Ca$. (b) and (d) drop mobility 
		versus time for the cases indicated in panel (a). The capillary number $Ca$ increases and 
		decreases periodically due to pinning-depinning events. (c) and (e) drop mobility for cases 
		indicated in panel (a) as a function of $Bo-Bo_c$. $Bo$ is the Bond number. The insets show 
		the critical Bond number, $Bo_c$, at which the drops start moving under external body force.}
	\label{fgr4:translational}
\end{figure*}
		
For large drop viscosity, viscous dissipation lies predominantly inside the drop. 
In this regime, as the drop viscosity is lowered, the drop velocity increases as 
$V \propto \eta_d^{-1}$ \cite{Keiser2017}, until it eventually plateaus to $V_x \simeq V_{ref}$. 
Both in simulations and experiments, the crossover occurs approximately at $\eta_d \sim 2\eta_l$. 
Below this value of drop viscosity, viscous dissipation in the lubricant ridge becomes dominant 
compared to dissipation in the drop. In this regime, the drop velocity has a strong dependence 
on the lubricant viscosity,  while the drop viscosity has virtually no effect.

There are a number of differences between the experiments in Ref.~\citenum{Keiser2017} 
and our numerical setup. Firstly, our simulations are in quasi 3-D, rather than full 3-D. Secondly, 
the size of the lubricant ridge compared to the drop size is larger than that in experiments. 
Thirdly, we have considered partial wetting lubricant, $\theta_{wet}=45^\circ$, whereas the experiments
were done using a complete wetting lubricant, $\theta_{wet}=0^\circ$.
Nonetheless, it is clear from Fig.~\ref{fgr3:viscous_dissipation} that the crossover between bulk drop and 
lubricant ridge dominated  dissipation regimes is a robust phenomenon, which our simulations 
can accurately capture.

\subsection{Variation in apparent and wetting angles}

In the previous section we numerically verified the crossover between viscous friction in 
the bulk drop and in the lubricant ridge. The similarity between experiments and numerical 
simulations is valid despite the fact we employed partial wetting lubricants, which involve 
also pinning and depinning effects. In this section we will focus on the role of lubricant 
wettability on the drop mobility, in particular on the interplay between contact line pinning 
and viscous friction.

As a reference case, we first consider a drop moving on a flat surface as illustrated in the top-left inset 
of Fig.~\ref{fgr4:translational}(a). The viscosity of the drop is set to be $\eta_d = 50 \eta_g$ to mimic a 
water drop in a dry air environment. A constant body force with $Bo=0.211$ is then applied to mobilise the 
drop so that the drop moves and reaches a steady state velocity. The results obtained for drops on a smooth surface 
are represented by black dots in Fig.~\ref{fgr4:translational}(a) as a function of the contact angle. 
For a smooth surface, we identify $\theta_{app} = \theta_{dg}^Y$.
In agreement with previous studies, the steady state capillary number of the drop increases monotonically 
with $\theta_{app}$ \cite{Moradi2011,Thampi2013}, due to the decrease in wedge dissipation at the contact line.

Let us now consider the equivalent setup for drops on LIS, as illustrated in the bottom-right inset of 
Fig.~\ref{fgr4:translational}(a). The lubricant viscosity $\eta_l$ is set to be the same as the drop viscosity, 
$\eta_l = \eta_d = 50 \eta_g$. To reduce the number of parameters to be explored in our simulations, 
we will assume a symmetric wetting condition for the lubricant, where $\theta^Y_{lg}=\theta^Y_{ld}=\theta_{wet}$.

For a given $\theta_{wet}$ we systematically vary the drop apparent angle, $\theta_{app}$, 
by tuning the fluid-fluid surface tensions, and consequently the Neumann angles, $\theta_{l}$, $\theta_{d}$ 
and $\theta_{g}$. In Fig.~\ref{fgr4:translational}(a) we compare the drop mobility, quantified as the time-averaged 
capillary number $Ca$ for $\theta_{wet} = 30^\circ$ (red triangles), $\theta_{wet} = 45^\circ$ (green diamonds), 
and $\theta_{wet} = 60^\circ$ (blue squares). 
Similar to the smooth surface case, the drop mobility increases monotonically with the apparent angle, 
but the magnitude of the $Ca$ is generally smaller than for a smooth surface.
Interestingly, when comparing the three datasets for different $\theta_{wet}$, we observe that, 
while for larger $\theta_{app}$ drops with smaller $\theta_{wet}$ move faster than those with larger $\theta_{wet}$, this ordering is reversed for lower $\theta_{app}$. The presence of these two regimes (for lower $\theta_{app}$ 
and larger $\theta_{app}$) is persistent for different values of $Bo$, $\eta_d$ and $\eta_l$.

We hypothesise this ordering inversion is due to a shift in the relative importance
between viscous dissipation and contact line pinning at the lubricant ridge.
To better characterise the pinning-depinning effects during drop motion, we plot the instantaneous 
$Ca$ associated to the drop's centre of mass, as a function of time for three drops with 
$\theta_{app} \sim 110^\circ$ and $\theta_{app} \sim 45^\circ$ respectively in Figs.~\ref{fgr4:translational}(b) 
and \ref{fgr4:translational}(d). 
We observe that the instantaneous $Ca$ oscillates periodically, which is due to pinning-depinning 
events as the drop moves across the periodic LIS pattern (see ESI video).
For both large (Fig.~\ref{fgr4:translational}(b)) and small (Fig.~\ref{fgr4:translational}(d)) $\theta_{app}$, 
the oscillations with larger amplitude are always observed for higher $\theta_{wet}$. 
At the same time, the amplitude of the oscillations is generally smaller for $\theta_{app} \sim 45^\circ$ than
for $\theta_{app} \sim 110^\circ$, which implies a less pronounced effect of pinning and depinning.

To further assess the relative importance of pinning versus viscous dissipation, we explore 
the relation between the driving force and the drop velocity for 
both cases of $\theta_{app} \sim 110^\circ$ (Fig.~\ref{fgr4:translational}(c)) 
and $\theta_{app} \sim 45^\circ$ (Fig.~\ref{fgr4:translational}(e)). Assuming a linear approximation, 
the relation between $Ca$, $Bo$ and $Bo_c$ can be expressed as $Ca = (Bo-Bo_c)/\beta$  
\cite{Smith2013, Varagnolo2013, Semprebon2014b}. $Bo_c$, the largest Bond number at which the 
drop remains stationary, is a measure of contact line pinning, or alternatively, contact angle hysteresis. 
$\beta$ is a function of the shapes of the drop and lubricant meniscus, and it is
related to their rate of viscous dissipation.

Considering $Ca$ as a function of $(Bo-Bo_c)$, our data show an important difference
between the large and small apparent angle drops. For large apparent angles (Fig.~\ref{fgr4:translational}(c)),
all the curves practically overlap onto a master curve. The variations in the results for 
$\theta_{wet} = 30^\circ$ (red triangles), $45^\circ$ (green diamonds), and $60^\circ$ (blue squares)
can be captured by differences in the value of the critical Bond number, $Bo_c$, as shown in the inset.
This indicates that the ordering observed in Fig.~\ref{fgr4:translational}(a) for large $\theta_{app}$
is determined by contact line pinning.
The prefactor $\beta$ is the same for the three datasets in Fig.~\ref{fgr4:translational}(c),
which suggest that the rate of viscous dissipation is on average the same once
the reduction in the effective driving force due to pinning forces is taken into account.

In contrast, for small apparent angles (Fig.~\ref{fgr4:translational}(e)), the datasets do not overlap 
onto a master curve. The critical Bond number, $Bo_c$, is also essentially the same -- 
any differences observed are within the error of the measurements -- for the three $\theta_{wet}$
used. These two observations suggest that, for low $\theta_{app}$, contact line pinning 
plays a minor role. The variations in $Ca$ vs $(Bo-Bo_c)$ for the three datasets
in $\theta_{wet}$ further imply that viscous dissipation is larger for the more wetting lubricant. 
Inspection of the drop morphologies supports this observation. We find that, for large $\theta_{app}$, the lubricant ridges have similar shape, regardless of $\theta_{wet}$. 
In contrast for low $\theta_{app}$ the ridge shape is broader for lower $\theta_{wet}$ 
(ESI document, SFig. 1 and SFig. 2).

\begin{figure}[t]
	\centering
	\includegraphics{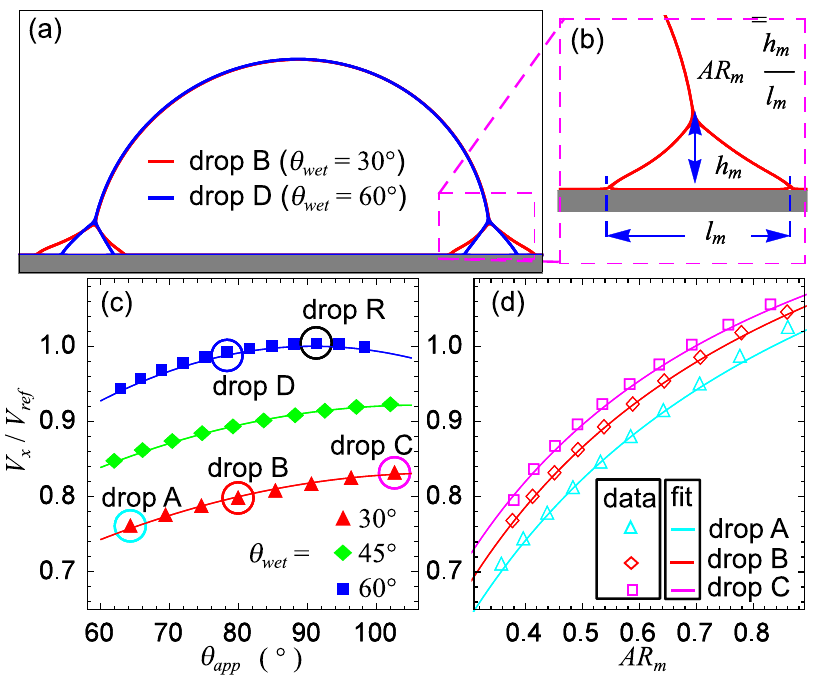}
	\caption{(a) In the absence of contact line pinning, drops with higher $\theta_{wet}$ 
	always move faster. (b) Comparison of drop shapes with the same $\theta_{app}$ but 
	different $\theta_{wet}$ and correspondingly meniscus aspect ratio $AR_m$. 
	(c) Definition of  $AR_m$. (d) Drop mobility against $AR_m$ for different $\theta_{app}$. 
	The lines are the best fit results to Eq.~\ref{equ:fitridge}. $V_{ref}$ is taken to be the velocity 
	of drop R in panel (a). 
	}
	\label{fgr5:aspect_ratio}
\end{figure}

To further corroborate this hypothesis, we ran three additional sets of simulations, 
where pinning and depinning is inhibited by replacing the topography with a flat
substrate, as shown in Fig.~\ref{fgr5:aspect_ratio}. The three sets correspond to $\theta_{wet} = 30^\circ$ 
(red triangles), $\theta_{wet} = 45^\circ$ (green diamonds), and $\theta_{wet} = 60^\circ$ (blue squares). 
The amount of lubricant in both the front and back ridge is the same for all cases.
Accordingly, once pinning is removed, drops with higher $\theta_{wet}$ always move faster 
irrespective of $\theta_{app}$, showing the same ordering that we obtain only for low 
$\theta_{app}$ in Fig.~\ref{fgr4:translational}(a).

Fig.~\ref{fgr5:aspect_ratio}(a) compares the morphologies of drops B and D indicated in Fig.~\ref{fgr5:aspect_ratio}(c). 
The two drops have an almost identical shape and $\theta_{app}$, but their lubricant ridge shapes and 
mobilities are different. For drop B, $\theta_{wet}$ is smaller, and therefore the meniscus is broader. 
We can characterise the meniscus shape by its aspect ratio, defined as $AR_m = h_m/l_m$, where $h_m$ and 
$l_m$ are its height and length respectively, see Fig.~\ref{fgr5:aspect_ratio}(b). 

We now propose a scaling argument to explain how the drop mobility depends on the lubricant ridge aspect ratio. 
We balance the rate of energy injected by the applied body force with the total rate of energy dissipation 
in the drop and lubricant,
\begin{equation}
F V_x \sim \eta_d \int |\nabla v|_d^2 dA_d + \eta_l \int |\nabla v|_l^2 dA_l.
\end{equation}
Here $F$ is the total force acting on the drop. We also recall that the simulations in Fig.~\ref{fgr5:aspect_ratio} 
are two-dimensional simulations; thus the terms on the right hand side are integrated over the drop and the 
lubricant ridge area. Taking $|\nabla v|_d \sim V_x/R$ and $|\nabla v|_l \sim V_x/h_m$ as the typical 
velocity gradient in the drop and lubricant meniscus, as well as $\Delta A_d \sim R^2$ and $\Delta A_l \sim h_m l_m$ 
as the typical scales for the cross-sectional area of the drop and the lubricant, we have 
\begin{eqnarray}
&F \sim \alpha_d \eta_d V_x + \alpha_l \eta_l V_x l_m/ h_m, \\
&V_x \sim \dfrac{F}{\alpha_d \eta_d + \alpha_l \eta_l / AR_m}. \label{equ:fitridge}
\end{eqnarray}
where $\alpha_d$ and $\alpha_l$ are positive, dimensionless fitting parameters.
Eq.~\ref{equ:fitridge} shows that a smaller $AR_m$ results in a larger energy dissipation 
in the lubricant meniscus, which in turn leads to the lower mobility of the drop. 

In Fig.~\ref{fgr5:aspect_ratio}(d), we consider drops A, B and C indicated in Fig.~\ref{fgr5:aspect_ratio}(c), 
and increase their $AR_m$ by tuning $\theta_{wet}$. We keep all other variables in the simulations 
the same, including the body force, the fluid surface tensions, the lubricant and drop viscosities, 
and the total drop and lubricant cross-sectional area. The data points in Fig.~\ref{fgr5:aspect_ratio}(d)
 correspond to simulation results, while the lines correspond to the best fit results to 
 Eq.~\ref{equ:fitridge}, where we have fitted $\alpha_d$ and $\alpha_l$ separately for 
each dataset. Consistent with our scaling argument, for all of the three datasets in Fig.~\ref{fgr5:aspect_ratio}(d),
drop mobility increases monotonically with $AR_m$.

Taking advantage of the results in Fig.~\ref{fgr5:aspect_ratio}, we can robustly conclude
that the ordering observed in Fig.~\ref{fgr4:translational}(a) for small $\theta_{app}$ is due to 
variations in viscous dissipation at the lubricant ridge. For the present 
choice of viscosities $\eta_l = \eta_d = 50 \eta_g$, the crossover between pinning and 
meniscus viscous friction dominated regimes in Fig.~\ref{fgr4:translational}(a)
occurs at  $\theta_{app}\simeq 70^\circ$.
In ESI SFig. 3, we take the limit where the lubricant viscosity is very low, equal to the gas viscosity.
In this case viscous dissipation at the lubricant is weak compared to that in the drop.
As expected, for low apparent angle $\theta_{app}$, we then observe that the drop mobilities remain very similar
as we vary the wetting angle $\theta_{wet}$.

\section{Conclusions}
\label{conclusion}
In this work we have employed a computational method, based on the free energy lattice Boltzmann 
approach, to study drop dynamics on LIS. We show that the drop apparent angle on LIS
can be captured accurately. Despite differences compared to typical experiments, namely the cylindrical 
geometry and the relatively larger size of the lubricant ridge, the drop mobility computed
from our simulations shows a remarkable agreement with the experiments by 
\citeauthor{Keiser2017}\cite{Keiser2017}, as the drop and lubricant viscosity ratio is varied.
Furthermore we have considered the more complex case of partially wetting lubricants,
and revealed a rich interplay between contact line pinning and viscous friction.
Specifically, we have shown that for large apparent angles contact line pinning 
dominates, and drops with more wetting lubricants move faster.
In contrast, for small apparent angles viscous friction in the lubricant ridge dominates.
The magnitude of the viscous dissipation is determined by the shape of the lubricant ridge,
and as such, drops in LIS with less wetting lubricants move faster.

To our best knowledge this is the first simulation study of drops on LIS that accounts for 
the full dynamics of the fluid flows.
The lattice Boltzmann method we have employed here is versatile, and there are a number of 
avenues of future numerical work. In this work we have assumed a LIS substrate textured with 
a regular periodic pattern of pillars, while many LIS substrates are constructed experimentally 
using irregular topographies \cite{Wong2011,Smith2013,Schellenberger2015,Luo2017}. 
The impact of random roughness on the drop dynamics will be investigated in a forthcoming study. 
It has also been pointed out that drainage of the infusing lubricant is a major source of failure 
for LIS technology \cite{Wexler2015,Kim2016}. As such, our approach is suitable for investigating
how the surface topographies can be designed to minimise the loss of lubricant during drop motion.

\section{Acknowledgement}
\label{Acknowledgement}
	
MSS is supported by an LPDP (Lembaga Pengelola Dana Pendidikan) scholarship from the Indonesian Government. HK acknowledges funding from EPSRC (grant EP/P007139/1) and Procter \& Gamble. CS acknowledges support from Northumbria University through the Vice-Chancellor's Fellowship Programme. We thank Matthew Wagner and Yonas Gizaw for useful discussions.

\bibliographystyle{apsrev4-1}
\bibliography{Ref}

\end{document}